\documentclass[conference]{IEEEtran}
\usepackage{hyperref}
\IEEEoverridecommandlockouts

\usepackage{cite}
\usepackage{amsmath,amssymb,amsfonts}
\usepackage{algorithmic}
\usepackage{graphicx}
\usepackage{textcomp}
\usepackage{xcolor}
\usepackage{makecell}
\usepackage{tabularx} 
\usepackage{algorithmic}
\usepackage{algorithm}

\def\BibTeX{{\rm B\kern-.05em{\sc i\kern-.025em b}\kern-.08em
    T\kern-.1667em\lower.7ex\hbox{E}\kern-.125emX}}
\begin{document}

\title{Near-Memory Architecture for Threshold-Ordinal Surface-Based Corner Detection of Event Cameras\\
}

\author{
    \IEEEauthorblockN{
        Hongyang Shang, 
        An Guo,
        Shuai Dong, 
        Junyi Yang,
        Ye Ke,
        Arindam Basu$^{\ast}$
    }
    \IEEEauthorblockA{
        Department of Electrical Engineering, City University of Hong Kong, Hong Kong, China\\
    }
    \thanks{$^{\ast}$Corresponding author (e-mail: arinbasu@cityu.edu.hk)

    This work was supported by the Research Grants Council of the HK SAR, China (Project No. CityU 11208125 and HKU C7003-24Y).
}
}

\maketitle

\begin{abstract}
Event-based Cameras (EBCs) are widely utilized in surveillance and autonomous driving applications due to their high speed and low power consumption. Corners are essential low-level features in event-driven computer vision, and novel algorithms utilizing event-based representations, such as Threshold-Ordinal Surface (TOS), have been developed for corner detection. However, the implementation of these algorithms on resource-constrained edge devices is hindered by significant latency, undermining the advantages of EBCs. To address this challenge, a near-memory architecture for efficient TOS updates (NM-TOS) is proposed. This architecture employs a read-write decoupled 8T SRAM cell and optimizes patch update speed through pipelining. Hardware-software co-optimized peripheral circuits and dynamic voltage and frequency scaling (DVFS) enable power and latency reductions. Compared to traditional digital implementations, our architecture reduces latency/energy by 24.7$\times$/1.2$\times$ at $V_{dd}=1.2$ V or 1.93$\times$/6.6$\times$ at $V_{dd}=0.6$ V based on 65nm CMOS process. Monte Carlo simulations confirm robust circuit operation, demonstrating zero bit error rate at operating voltages above 0.62 V, with only 0.2\% at 0.61 V and 2.5\% at 0.6 V. Corner detection evaluation using precision-recall area under curve (AUC) metrics reveals minor AUC reductions of 0.027 and 0.015 at 0.6 V for two popular EBC datasets.

\end{abstract}

\begin{IEEEkeywords}
Near-Memory Computing, Event Camera, Corner Detection, Threshold-Ordinal Surface.
\end{IEEEkeywords}

\section{Introduction}
\label{sec:intro}

Event-based cameras (EBCs), also known as dynamic vision sensors (DVS), have garnered significant attention due to their exceptional low latency performance and low power consumption. Unlike conventional cameras that capture frames at fixed intervals, event cameras detect changes in contrast over time. When a pixel’s contrast change exceeds a threshold, an event is triggered and  an asynchronous output- the pixel’s position, the timestamp, and the polarity (whether the brightness increased or decreased) is transmitted. As a result, EBCs provide higher dynamic range, lower latency, and reduced bandwidth compared to conventional CMOS image sensors (CIS) \cite{gallego2020event}. They are capable of achieving effective frame rates exceeding $>$ 5000 fps \cite{rebecq2019high}, while conventional video cameras are typically limited to 100 fps. 

This unique capability of EBCs makes them an excellent choice for real-time applications such as autonomous driving \cite{chen2020event}, robotics \cite{mahlknecht2022exploring}, and unmanned aerial vehicles (UAVs) \cite{shariff2024event}, which rely on fast computer vision. In computer vision, corner detection is a fundamental processing step, playing a vital role in high-level applications such as object tracking \cite{du2020correlation}, 3D reconstruction \cite{reddy2018carfusion}, and object recognition \cite{bansal2021efficient}. It helps identify key features within an image or video stream that are stable and distinguishable from their surroundings. 

\begin{figure}[t]
\centerline{\includegraphics[width=0.5\textwidth]{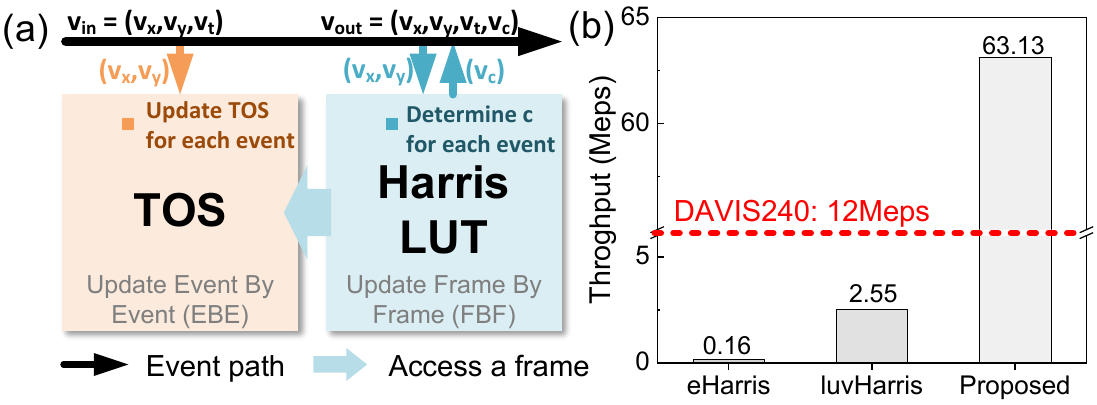}}
\caption{(a) Architecture of luvHarris where the TOS is updated for each incoming event, $v_{in}$. The Harris lookup table (LUT) is updated for the full frame by accessing the TOS and $v_{in}$ is tagged as a corner $c$ or not by referencing the last available Harris LUT.(b) The maximum throughput of eHarris, the conventional implementation of luvHarris, and the proposed NMC-TOS compared with the maximum bandwidth of the DAVIS240\cite{gallego2020event}. }
\label{fig:motivation}
\end{figure}

However, traditional corner detection algorithms are designed for frames. Modified versions have been proposed for EBCs, but they exhibit significant limitations when processing EBC event streams. The eHarris approach \cite{vasco2016fast} retains the event by event (EBE) processing nature, achieves decent accuracy, but imposes prohibitive computational overhead (and concomitantly latency--see \cite{glover2021luvharris} and Fig. \ref{fig:motivation}) by requiring Harris score calculations for every asynchronous event. While segment-based methods such as FAST \cite {mueggler2017fast} and ARC \cite{alzugaray2018asynchronous} improve throughput by identifying continuous segments
of events, they suffer from elevated false positive rates in cluttered environments due to inherent noise sensitivity. The luvHarris algorithm \cite{glover2021luvharris} addresses these issues by decoupling the Harris computation via an as fast as possible, EBE update of Threshold-Ordinal Surface (TOS) and a separate frame by frame(FBF) Harris update that uses the last available TOS, as shown in Fig.\ref{fig:motivation}(a). However, the increasing event rate from high-resolution EBCs still gets constrained in the TOS update mechanism, as shown next.

Updating a TOS patch of $P\times P$ using traditional digital circuit implementation leads to O($P^2$) computational complexity. For a 7$\times$7 patch size and a 500 MHz clock frequency, the event update latency is around 392 ns, yielding an event processing rate of only $\approx$2.6 Meps—much below the bandwidth of most current event cameras \cite{gallego2020event} (as shown in Fig. \ref{fig:motivation}(b)), leading to event loss at higher event frequencies. The Harris score calculation is not a problem for the high resolution EBCs since the major operation is a convolution which can be efficiently accelerated by the modern in-memory chips designed for convolutional neural networks (CNN). For example, Harris score calculation for a 1280$\times$720 resolution with 5x5 Sobel and Harris windows requires $\approx$236 Mops. Given the high throughput of many recent CNN chips \cite{zimmer20200,matsubara202112}, it is estimated that Harris LUT update rate $>$ 1 KHz. Therefore, although intuitively the Harris score calculation might seem more complex, the per-event update requirement of TOS makes it the bottleneck for hardware implementation, particularly for the high event rate of modern resolution event cameras \cite{sony_prophesee_imx636}.

To address this issue, we propose a near-memory architecture designed for efficient TOS generation following the recent trend of using Near/In-Memory Computing for event cameras \cite{bose2022389,xueyong}. As shown in Fig. \ref{fig:motivation}(b), this can push the supported event rate to $\approx$ 63 Meps, compatible with high resolution EBCs. The main contributions of this work are:
\begin{itemize}
    \item A near-memory computing architecture (NMC-TOS) for the TOS update with simplified minus one logic (MOL) and comparison  modules (CMP) for reduced delay and power.  
    \item A read-write decoupled 8T SRAM with pipelining approach to further reduce the delay required for processing a patch.
    \item Dynamic voltage and frequency scaling (DVFS) to adjust the chip's voltage dynamically based on the event rate for reduced average power consumption.
    \item Using the above techniques, we demonstrate a speedup of 1.9$\times$ to 24.7$\times$ and reduced power consumption by a factor of 1.2$\times$ to 6.6$\times$ compared to the traditional implementation.
    \item System level simulations at the worst-case operating voltage of 0.6 V show a decrease of 0.027 and 0.015 in the area under the precision-recall curve (AUC) for the shapes\textunderscore dof\cite{mueggler2017event} and dynamic\textunderscore dof\cite{mueggler2017event} datasets, respectively.

\end{itemize}
 
\section{BACKGROUND}
\label{sec:background}
\subsection{Event Based Camera and Data Representation}
Event cameras generate sparse event outputs that encode changes in temporal contrast, making them highly effective for high-speed, real-time applications. Each event, $v=(v_x,v_y,v_p,v_t)$ is typically represented using the Address Event Representation (AER) protocol where the coordinates ($v_x$ and $v_y$), the polarity ($v_p$), and a timestamp ($v_t$) are used. The polarity indicates whether the event is triggered by an increase (ON event) or decrease (OFF event) in temporal contrast while the timestamp records the triggering time of the event. 

\subsection{Corner Detection Methods for EBC}
Since event camera data are in AER format rather than the traditional 2D image format, some algorithms have been developed specifically for corner detection based on event data. 

The eHarris algorithm\cite{vasco2016fast} creates a binary surface of active events indicating the occurrence of an event in the recent past. The corner is computed by evaluating the Harris score of the binary surface for each event. Although the accuracy of corner classification using the eHarris method achieves higher accuracy than ARC\cite{alzugaray2018asynchronous} \& FAST\cite{mueggler2017fast}, their computational complexity results in high latency since the Harris calculation has to be performed for every event.

Segment-based algorithms like FAST\cite{mueggler2017fast} and ARC\cite{alzugaray2018asynchronous} can be directly applied to the Surface of Active Events (SAE), performing calculations at a higher throughput because they do not require complex operations on the whole image. These algorithms detects corners by comparing the event timestamps in the local neighborhood around the current event. They use a circular neighborhood mask to identify continuous segments of events with higher timestamps than the surrounding pixels. If these segments form a continuous arc, the current event is classified as a corner in \cite{mueggler2017fast} while if the angle of this arc exceeds a certain threshold (usually 90 degrees), the event is classified as a corner in \cite{alzugaray2018asynchronous}. However, both FAST and ARC only utilize information from the local circular region, making them more sensitive to noise. This sensitivity can lead to high false positive rates\cite{glover2021luvharris}.

The luvHarris algorithm\cite{glover2021luvharris} aims to improve corner detection throughput while maintaining high accuracy. To achieve this, luvHarris introduces two key innovations: first, it proposes a variant of the SAE called the Threshold-Ordinal Surface (TOS), which uses 8-bit unsigned integers to represent the novelty of the recorded events. Second, luvHarris decouples the event-by-event TOS update from the frame-based Harris score calculation, and uses the last computed Harris score as a look up table (LUT). However, the method is demonstrated on high-end, multithreaded processors and not suited for edge applications. Simpler processors can achieve limited throughput due to the TOS update per event as shown in Section \ref{sec:intro}.

\section{Overall Architecture}
\label{sec:architecture}
\begin{figure}[t]
\centerline{\includegraphics[width=0.4\textwidth]{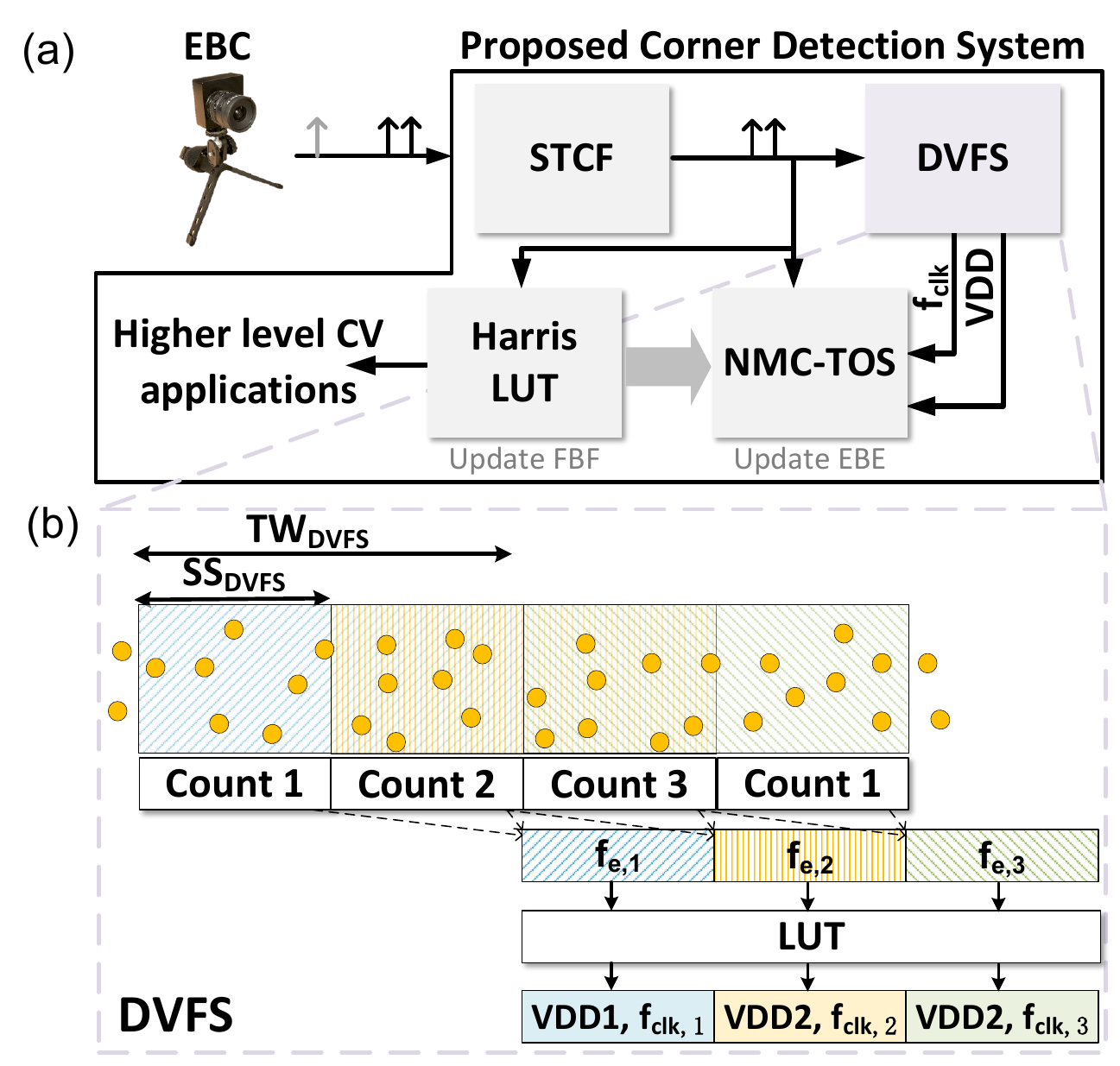}}
\caption{The workflow for corner detection on the event stream of an EBC. The events received by the EBC first pass through an STCF filter to remove noise. Then, the TOS is constructed by NMC-TOS EBE, while the event frequency is detected using the DVFS module. Finally, corner detection is performed FBF.}
\label{fig1-overview}
\end{figure}
Fig. \ref{fig1-overview}(a) illustrates the overview of the corner detection system, which includes the blocks of Spatio-temporal Correlation Filtering (STCF), DVFS, TOS and corner detector.

\subsection{Spatio-Temporal Correlation Filtering Denoising}\label{ITH}

The Spatio-Temporal Correlation Filtering (STCF) method is designed to reduce background activity (BA) noise in EBCs\cite{guo2022low}. BA noise events are typically isolated, while signal events occur in groups of temporally and spatially correlated events. The STCF filter checks the event count within a spatio-temporal neighborhood defined by a time window (TW$_{STCF}$). If enough supporting events (e.g., 2) are present, it is classified as a signal, else, it is classified as noise.

\subsection{Dynamic Voltage and Frequency Scaling}\label{FAT}

Since event cameras capture only temporal contrast changes per pixel which is dependent on the scene, the event rate fluctuates rather than being fixed like the frame rate of traditional cameras. This has been historically exploited by designing asynchronous hardware \cite{akopyan2015truenorth, davies2018loihi}, but the lack of publicly available asynchronous design tools makes its adoption difficult. Instead, we propose to dynamically adjust the clock frequency based on the event rate, a method compatible with traditional synchronous design. By using lower voltage during low event rates and adjusting the clock to a lower frequency, the energy consumption for each event when updating the TOS is effectively reduced. As shown in Fig. \ref{fig1-overview}(b), a moving window averaging is employed to count the number of events with window size $TW_{DVFS}$ and stride $SS_{DVFS}$ (we fix stride to be $50\%$ in this work). The value of $TW_{DVFS}$ and the bitwidth of the counters depend on application; for driving datasets\cite{prophesee_datasets}, $TW_{DVFS}=10$ ms and bitwidth of 20 were sufficient.

A round-robin counter mechanism is employed to perform the counting operation (Fig. \ref{fig1-overview}(b)). The method involves three counters working in sequence, such that when one counter is counting, the other two are used to provide an estimate of the event rate $f_{e}$. As shown in Fig. \ref{fig1-overview}(b), each counter's duration is fixed to $\frac{TW_{DVFS}}{2}$. A pointer $ptr$ determines the current counter accepting events and its position is shifted among the three counters in a circular sequence given by: $ptr\leftarrow (ptr+1)mod(3)$. The eventf frequency $f_{e}$ is mapped to the required dynamic voltage($VDD$) and clock frequency ($f_{clk}$) via a LUT.


\subsection{NMC-TOS and Harris Based Corner Detector}\label{sec:FAT}

\begin{algorithm}[!t]
\caption{Event-by-event TOS update}
\label{alg:alg1}
\begin{algorithmic}
  \STATE \textbf{Require:} $v = \langle v_x,v_y,v_t\rangle$, $TOS$
  \STATE \textbf{for} $x = v_x - \tfrac{P-1}{2} : v_x + \tfrac{P-1}{2}$
  \STATE \hspace{0.5cm}\textbf{for} $y = v_y - \tfrac{P-1}{2} : v_y + \tfrac{P-1}{2}$
  \STATE \hspace{1.0cm}$TOS_{xy} \gets TOS_{xy}-1$
  \STATE \hspace{1.0cm}\textbf{if} $TOS_{xy} < TH$ \textbf{then} $TOS_{xy} \gets 0$
  \STATE $TOS_{v_xv_y} \gets 255$
\end{algorithmic}
\end{algorithm}

TOS uses 8-bit unsigned integers to represent the novelty of events. Details of the algorithm are shown in Algorithm \ref{alg:alg1}. Whenever a new event $v$ occurs, the TOS updates the corresponding patch data by subtracting 1 (where the patch size is $P$) and applies a threshold with a fixed value, TH. Then, the pixel location is updated to 255. Hardware implementation of TOS using near memory computing (NMC-TOS) is illustrated in section \ref{sec:NMCTOS}. The standard Harris operator in run on this TOS representation to detect corners\cite{glover2021luvharris}.

\section{Near-Memory Computing for TOS}
\label{sec:NMCTOS}

\begin{figure}[t]
\centerline{\includegraphics[width=0.4\textwidth]{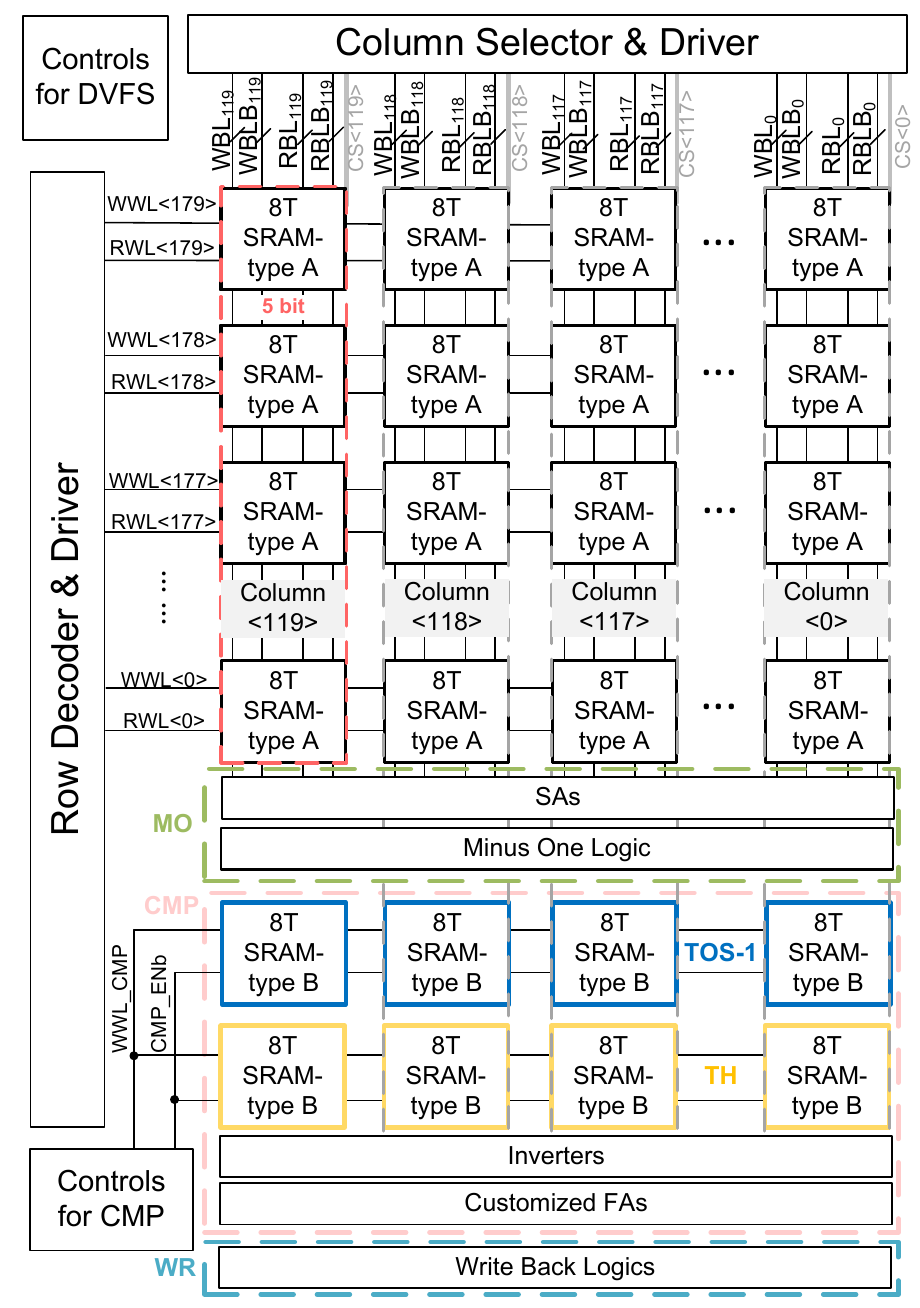}}
\caption{The overall block architecture of NMC-TOS is divided; an EBC like DAVIS240 with resolution $240\times180$ requires two such blocks. Each block of the TOS array consists of 180 rows and 120 columns of 5-bit words. The peripheral circuits include MO module, CMP module, WR module, buffer, and control circuits.}
\label{fig:arch}
\end{figure}

\subsection{Hardware Overview}
\label{sec:hw_overview}
To optimize the TOS memory size, we note that although the maximum TOS value is 255 requiring 8 bits for storage, the threshold typically does not go below $\approx$225 in practice. As a result, the highest 3 bits can be omitted from on-chip storage to save space, and each TOS is stored using only 5 bit words. The architecture uses a block size of 180$\times$600 8T SRAM (type A) array, as shown in Fig. \ref{fig:arch}, which corresponds to storing 5-bit words for a sensor resolution of 180$\times$120. This block size can be repeated as many times as needed to accommodate different resolution cameras. For example, two such blocks are required to cover the entire sensor for a DAVIS240 sensor whose resolution is 240$\times$ 180. The overall architecture of a single block, as shown in Fig. \ref{fig:arch}, includes the DVFS control module, row/column selectors, drivers, a 180$\times$600 8T SRAM (type A) array, the minus one (MO) module, the compare (CMP) module, and the write-back (WR) module. The MO module consists of the sense amplifier (SA) and the minus one logic (MO), while the CMP module includes the 2$\times$600 8T SRAM (type B), inverters, and the customized full adder (FA).


\begin{figure}[t]
\centerline{\includegraphics[width=0.5\textwidth]{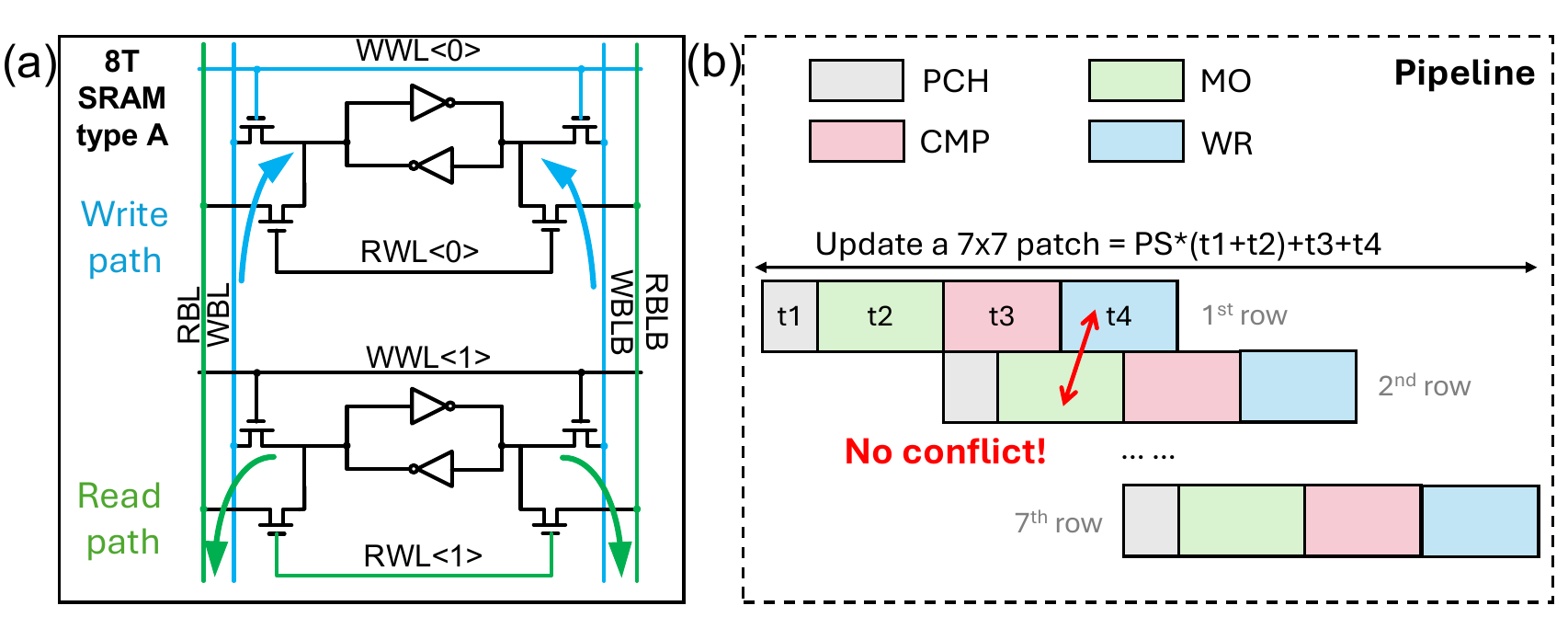}}
\caption{8T SRAM-type A and the pipeline method. (a) Decoupling the WBL and RBL makes it possible for the write-back and read operations to occur simultaneously. (b) A pipeline example for updating a 7$\times$7 patch.}
\label{fig3_pipeline}
\end{figure}

\subsection{8T SRAM for Pipeline}
\label{sec:pipeline}
For the SRAM array that stores the TOS value, a read-write decoupled 8T SRAM architecture is employed, as shown in Fig. \ref{fig3_pipeline}(a). This SRAM structure separates the read bitline (RBL) and write bitline (WBL) by replicating two access NMOS transistors, allowing reading and writing to occur simultaneously. Assuming the patch size is $P$, without this architecture, the reading of the next row must occur after the write-back of the previous row, which introduces a delay of $(t1 + t2 + t3 + t4) \times P$ where the $t_i$ correspond to the delays of four phases of operation as shown in Fig. \ref{fig3_pipeline}(b). However, with the 8T SRAM architecture, pipelining can be used for the multiple computation steps, compressing the required computational delay to $P \times (t1 + t2) + t3 + t4$, which decreases the delay by about 2$\times$.

\begin{figure}[t]
\centerline{\includegraphics[width=0.45\textwidth]{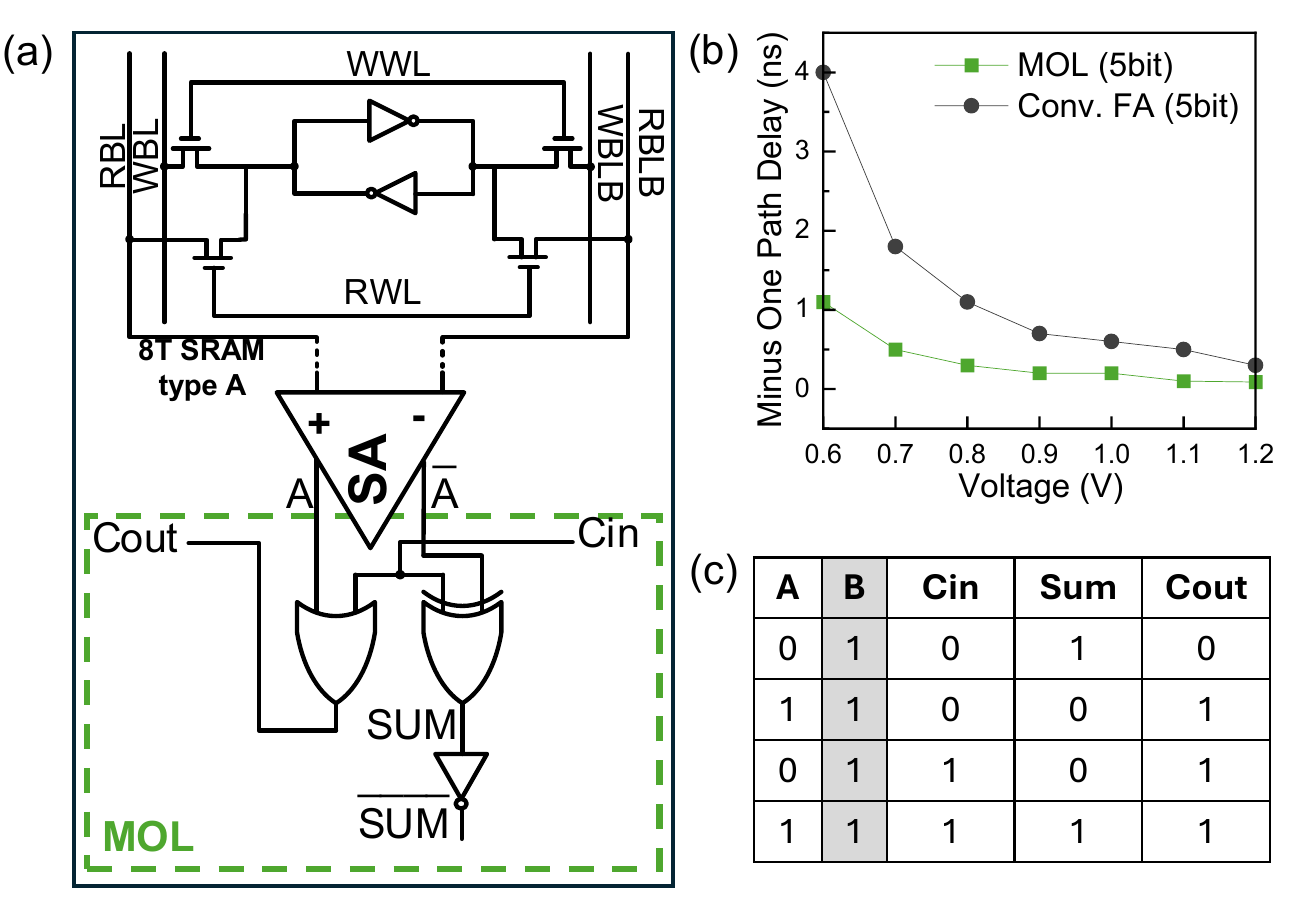}}
\caption{Minus one (MO) module. (a) MO module SA for readout of SRAM and simplified minus one logic (MOL). (b) MOL reduces path delay compared to 28T full adders (FA). (c) Truth table of MOL.}
\label{MOL}
\end{figure}

\subsection{Minus One (MO) Module and Comparison (CMP) Module}
\label{sec:mol_cmp}
For the decrement by 1 operation in neighbouring pixels, the MO module is shown in Fig. \ref{MOL}(a). The TOS stored in 8T SRAM type A is read out by a latched SA \cite{yu2020logic}, and directly sent to the simplified minus one logic (MOL). Compared to the conventional 28T full adder, the computational delay is smaller, as shown in \ref{MOL}(b). Since one addend B is always -1 (all bits are 1 in 2's complement), the full adder is simplified to avoid the logical redundancy typically introduced by using a 28T full adder in traditional designs \cite{guo202428}, as shown in the truth table of Fig. \ref{MOL}(c). 

\begin{figure}[t]
\centerline{\includegraphics[width=0.45\textwidth]{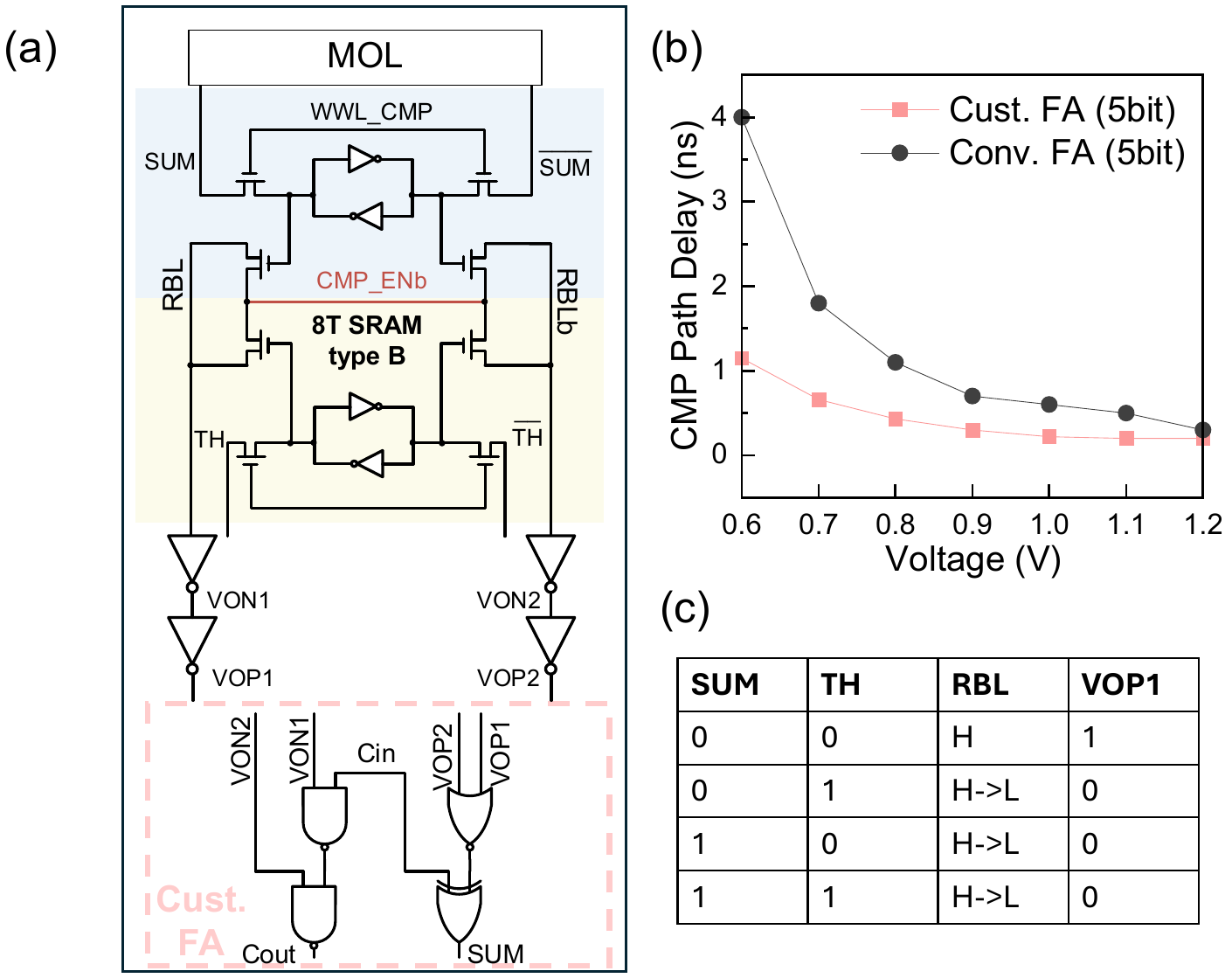}}
\caption{Comparison (CMP) module. (a) CMP consists of two rows of 8T SRAM type B (one stores the MOL result and the other stores threshold (TH)), inverters for readout and customized FAs. (b) Customized FA reduces path delay compared to 28T FA. (c) Operation principle of CMP.}
\label{CMP}
\end{figure}

For the CMP module, another type of 8T SRAM is adopted as shown in Fig. \ref{CMP}(a), with one row used to store the SUM output of the MOL (which is TOS-1) and another row to store the threshold (TH). The computation method of this architecture is an NOR operation (as shown in Fig. \ref{CMP}(c)), meaning that the RBL  only remains high if both of SUM and TH at the same bit significance are stored as 0; otherwise, RBL drops to 0 V. Notice that, RBL in the CMP module is decoupled from the big capacitance of SRAM array's RBL. As a benefit, this SRAM allows RBL to achieve full swing with only two rows for computation. Hence, inverters can be used to determine the status of RBL and RBLB, significantly reducing power consumption compared to SA. Based on the characteristics of this NOR computation, the subsequent custom full adder is implemented instead of a traditional full adder \cite{chen2023bp}, which further reduces the delay as demonstrated in Fig. \ref{CMP}(b).

\subsection{Detailed operation of the NMC Architecture}
\begin{figure}[t]
\centerline{\includegraphics[width=0.4\textwidth]{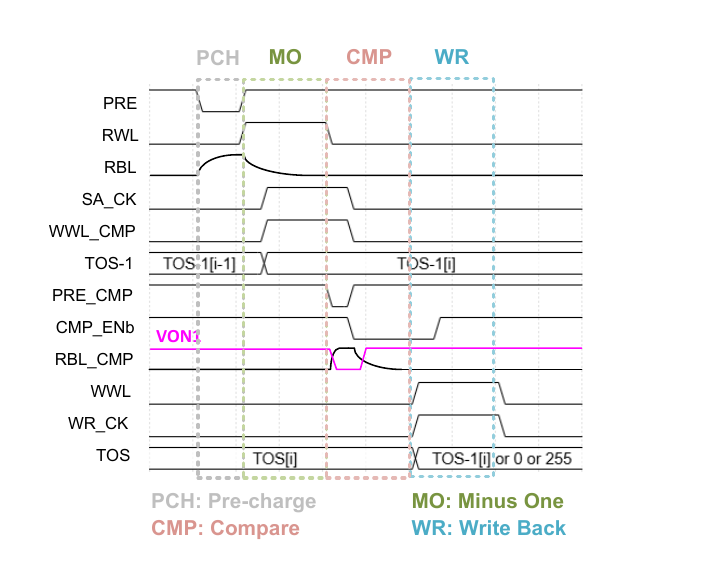}}
\caption{The overall timing diagram of one row calculation in NMC-TOS comprising precharge (PCH), minus-one (MO), comparison (CMP) and write back (WR).}
\label{all_wave}
\end{figure}

The waveform of the NMC-TOS is shown in Fig. \ref{all_wave}. For different operating voltage given by the DFVS module, the required clock cycles of the four phases are same, but the clock frequency is different.
As described in Sec. \ref{sec:pipeline}, the NMC-TOS primarily involves four working stages: PCH, MO, CMP, and WR. These stages are performed sequentially for each row in the patch. Initially, the PRE signal triggers a low pulse to precharge the RBL of the SRAM-type A  array, preparing for the subsequent reading of the stored TOS. Then, during the MO stage, the RWL signal rises, and if the TOS stores a 0, the RBL will decrease. At this point, the SA clock (SA\textunderscore CK) arrives slightly later to ensure setup time. Meanwhile, the WWL\textunderscore CMP is enabled, allowing the result of the MO to be written to the TOS-1 SRAM in the CMP module. During this phase, the precharge of RBL\textunderscore CMP (PRE\textunderscore CMP) is also occurring for the next calculation. In the next CMP stage, the arrival of CMP\textunderscore ENb causes the voltage on RBL\textunderscore CMP to change, and VON/VOP is output accordingly (as shown by the pink line in the figure). Finally, in the WR stage, WR\textunderscore CK controls the DFF to latch the write back value (TOS-1 or 0 or 255) and WWL is enabled to write the value back into the main TOS SRAM. The value latched by the DFF is determined by the Cout signal from the most significant bit of the CMP module.

\section{Results}
\label{sec:results}
The NMC macro was implemented in a 65 nm CMOS process and SPICE simulation results are reported next.
\subsection{Dynamic Voltage Frequency Scaling}
\label{sec:result_dvfs}
\begin{figure}[t]
\centerline{\includegraphics[width=0.45\textwidth]{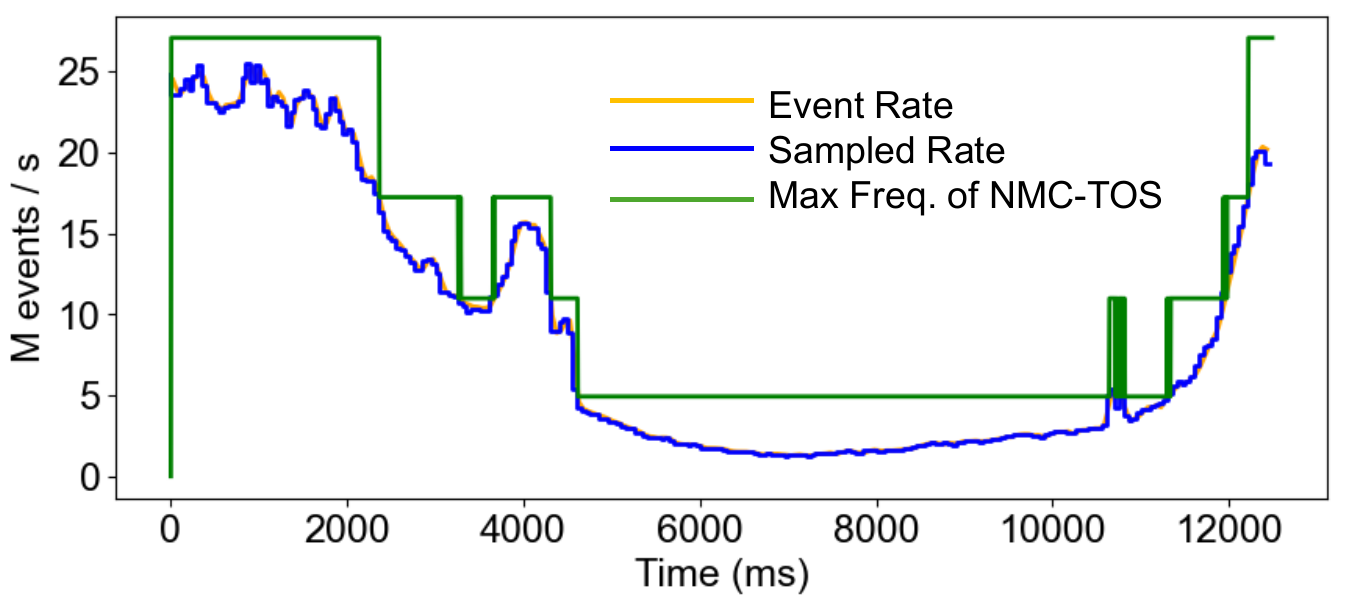}}
\caption{The effect of DVFS in the driving dataset \cite{prophesee_datasets} shows the proposed method can modify the maximum event handling capacity according to fluctuations in event rate.}
\label{DVFS}
\end{figure}

The effect of DVFS is shown in Fig. \ref{DVFS} for 111.4M number of events from the driving dataset\cite{prophesee_datasets}.
It can be observed that the event rate sampled by the 20-bit counter, sensitively captures changes in the event rate. Additionally, the DVFS module dynamically adjusts the operating voltage and maximum event frequency of the NMC-TOS based on the sampled rate. Throughout the entire dataset, the event rate never reached the maximum operating frequency of 63.1 Meps at 1.2V, indicating that the NMC-TOS can continuously update the TOS without any event loss during operation in this dataset. Table \ref{DVFS_table} summarizes the event rates and corresponding power savings for different datasets\cite{prophesee_datasets,mueggler2017event} with 1.4$\times$ to 5.3$\times$ savings achievable.



\begin{table}[t]
\centering
\caption{Power improvement using DVFS under several datasets}
\begin{tabular}{c c c c c c}
\hline
\makecell[c]{Dataset} & \makecell[c]{Max Event \\ Rate (Meps)} & \makecell[c]{\# Events \\(M)} & \makecell[c]{Power w\\ DVFS (mW)} & \makecell[c]{Power w/o \\ DVFS (mW)}\\ \hline
driving\cite{prophesee_datasets} & 25.9 & 111.4 & 0.44 & 1.24\\  
laser\cite{prophesee_datasets} & 39.5&57.6 & 3.90& 5.37\\ 
spinner\cite{prophesee_datasets} &11.4 & 54.1& 0.38& 1.50\\ 
dynamic\textunderscore dof\cite{mueggler2017event} &4.5 & 57.1&0.02 & 0.13\\ 
shapes\textunderscore dof\cite{mueggler2017event} &1.9 &18.0 &0.01 &0.04 \\ \hline
\end{tabular}
\label{DVFS_table}
\end{table}

\subsection{Latency and Energy Comparisons}

\begin{figure}[t]
\centerline{\includegraphics[width=0.5\textwidth]{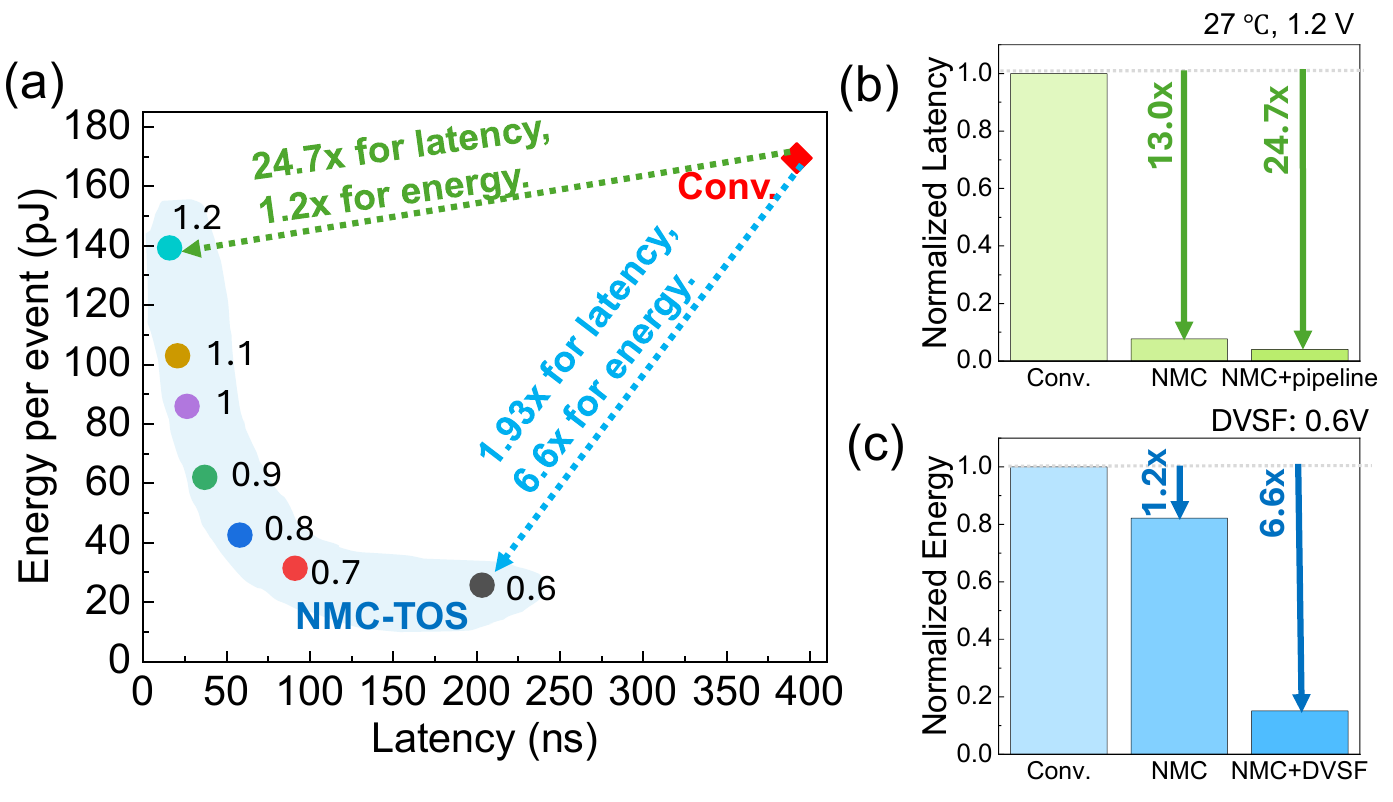}}
\caption{(a) Energy consumption and latency of the conventional digital circuit method and the NMC-TOS architecture at different operating voltages. (b) Impact of using NMC method and NMC+pipeline on latency. (c) Impact of using NMC method and NMC+DVFS on energy consumption.
}\label{result1}
\end{figure}

As shown in Fig. \ref{result1}(a), the blue region represents the results of NMC-TOS for updating a patch, considering the effects of pipeline and the varied operating voltage from 1.2 V to 0.6 V. It can be observed that at an operating voltage of 1.2 V, the latency is the lowest (16 ns), while the energy consumption is relatively high (139 pJ). As the operating voltage decreases to 0.6 V, latency increases to 203 ns, but energy consumption decreases to 26 pJ. From  Fig. \ref{result1}(b), it can be seen that using the NMC architecture reduces the latency by 13.0$\times$ compared to the conventional method, and employing a pipeline further reduces the latency by 24.7$\times$. Regarding the normalized energy (as shown in Fig.\ref{result1}(c)), the NMC architecture reduces energy consumption by 1.2$\times$ compared to the conventional method. Moreover, using DVFS at the most aggressive setting (operating at 0.6 V) allows for a further reduction in energy consumption by 6.6$\times$.

\begin{figure}[t]
\centerline{\includegraphics[width=0.45\textwidth]{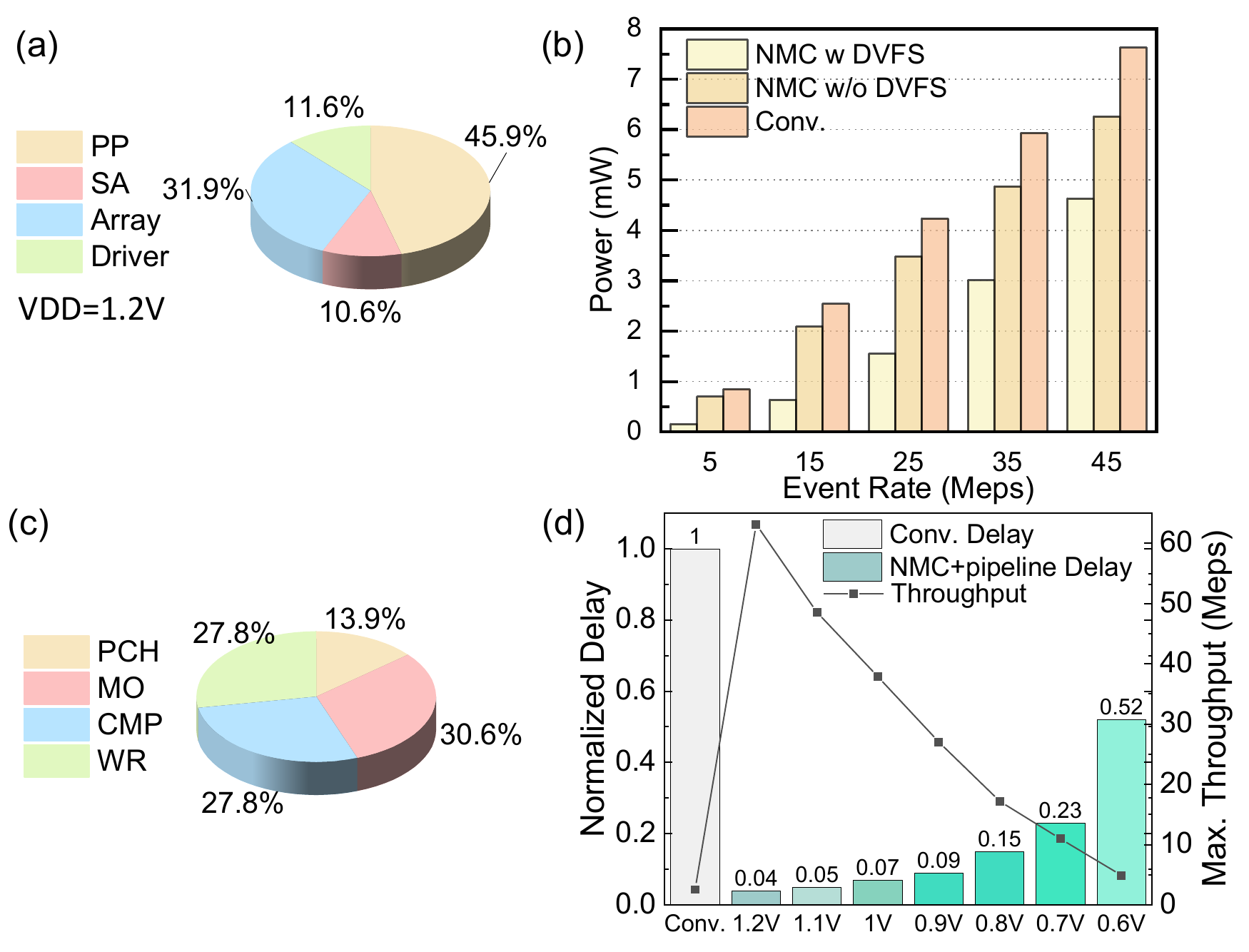}}
\caption{Analysis of power, delay and throughput in NMC-TOS. (a) Breakdown of energy consumption for each module at an operating voltage of 1.2V. (b) Comparison of power consumption at different event frequencies. (c) Delay of different computation phases at an operating voltage of 0.6V. (d) Comparison of computational latency per event and maximum throughput that can be handled by conventional methods and NMC-TOS. 
}
\label{result_power_delay}
\end{figure}

Detailed power breakdown shown in Fig. \ref{result_power_delay}(a) indicates that the peripheral circuits (PP) is the main source of power consumption, accounting for 45.9\%, which is due to the placement of both the computation and write-back modules in the peripheral circuits. Additionally, the array account for 31.9\%. The driver and the SA used for SRAM array reads contribute 11.6\% and 10.6\%, respectively. For different event rates, it can be observed from Fig. \ref{result_power_delay}(b) that the NMC method consistently outperforms the conventional method, and the use of DVFS further reduces power consumption. For example, at the event rate of 45 Meps, NMC reduces power dissipation by 1.2$\times$ compared to the conventional one while adding DVFS to it provides a further reduction of 1.37$\times$. 

In terms of delay, it is evident in Fig.\ref{result_power_delay}(c) that the minus one (MO) phase exhibits the largest delay, accounting for 30.6\%. The delays for the other phases are as follows: pre-charge (PCH) accounts for 13.9\%, computation (CMP) accounts for 27.8\%, and write-back (WR) accounts for 27.8\%. As shown in Fig. \ref{result_power_delay}(d), it is observed that at higher operating voltages, the computation delay decreases and throughput increases. Compared to the conventional method, updating a 7$\times$7 TOS patch benefits from the parallelism of the NMC architecture, and the addition of a pipeline further reduces the delay. Compared to the traditional method's throughput of 2.6 Meps, NMC-TOS achieves a throughput of 63.1-4.9 Meps. Even at the lowest operating voltage of 0.6V, the throughput is improved by 1.9$\times$ over the conventional one.


\subsection{Impact of Hardware Non-Ideality on the Algorithm}
\label{sec:hw_error}
\begin{figure}[t]
\centerline{\includegraphics[width=0.45\textwidth]{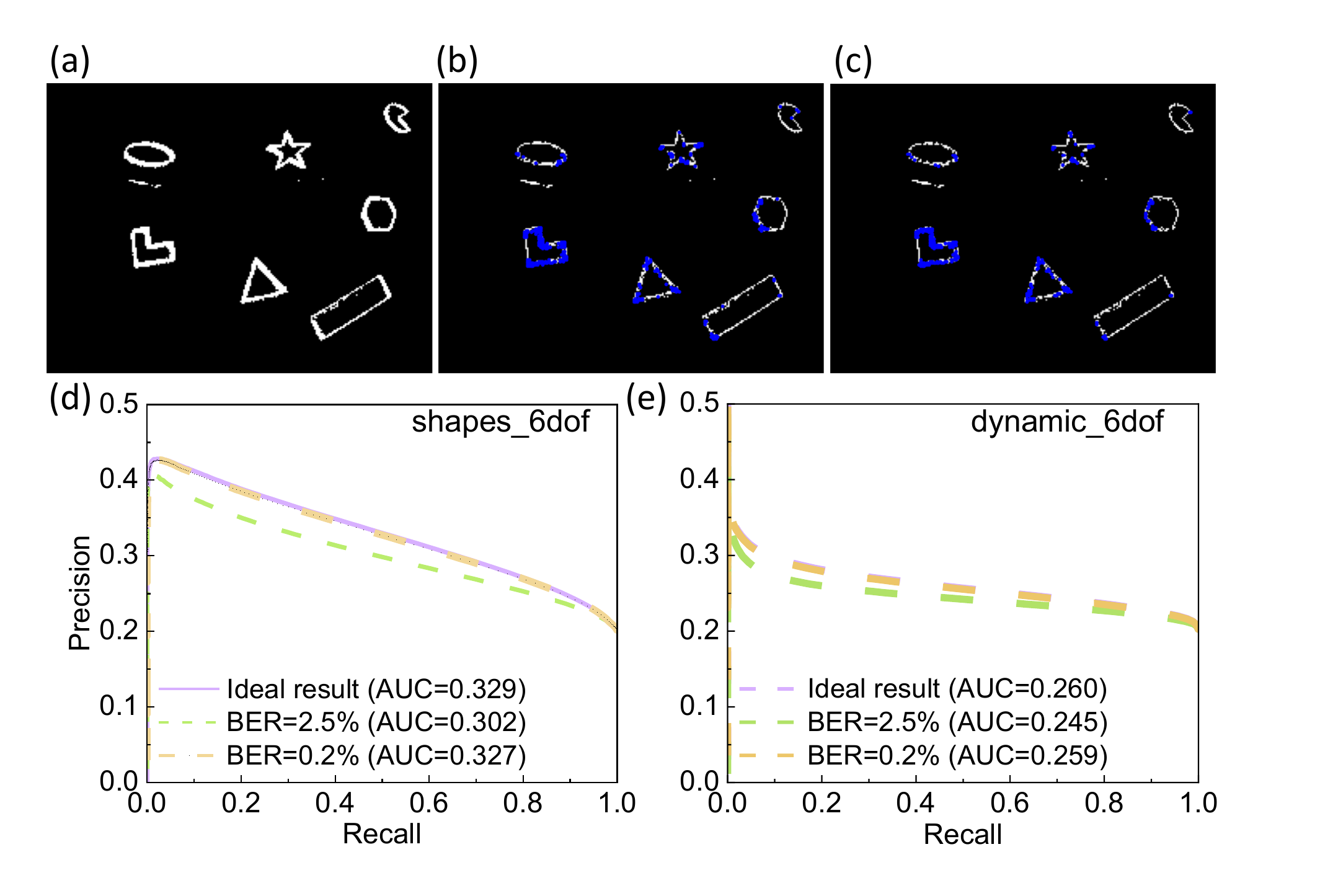}}
\caption{Visualizing 5000 events from the shapes\textunderscore dof dataset. (a) SAE representation of timestamps. (b) TOS without precision loss and the detected corner. (c) TOS considering BER at 0.6V and the detected corner. (d) Precision-Recall results on the (d) shapes\textunderscore dof and (e) dynamic\textunderscore dof.}
\label{img_quality_pr_curve}
\end{figure}

To estimate errors due to hardware non-idealities, Monte Carlo simulations were run to estimate BER which showed errors only occur below 0.62 V, with a BER of 0.2\% at 0.61V and 2.5\% at 0.6 V. In the write-back circuit, when the value stored in the original TOS memory is 0, the write-back is disabled, meaning that the error occurs only in pixels with valid values. 
Additionally, since the highest 3 bits are omitted as mentioned in \ref{sec:hw_overview}, the erroneous values range between 224 and 255, significantly reducing the impact on the computation. We analyzed the effect of the worst-case BER on the computation by injecting errors in a software simulation of the pipeline on shapes\textunderscore dof dataset \cite{mueggler2017event}. Fig. \ref{img_quality_pr_curve} compares SAE \cite{mueggler2017fast} using timestamps as grayscale values in (a), error-free TOS with corner results in (b), TOS considering hardware errors with corner results in (c). It can be observed that the hardware errors do not significantly affect the results visually.

To quantitatively assess the impact of errors, we compared precision-recall curves for the corner detection results as done in \cite{glover2021luvharris}. As shown in Fig.\ref{img_quality_pr_curve}(c) and (d), it can be observed that, on the shapes\textunderscore dof dataset, when the BER is 2.5\%, the AUC decreases by only about 0.027 and remains unchanged when the BER is 0.2\%. For the dynamic\textunderscore dof dataset, the AUC decreased by only 0.015 at a BER of 2.5\% and is again almost unchanged at a BER of 0.2\%. This shows possibility of using DVFS with power supplies down to 0.6 V in our pipeline.


\section{Conclusion}
This work addresses the issues of low throughput and high power consumption in state-of-the-art corner detection algorithms for event cameras using TOS. Implementation of earlier hardware architectures in edge devices result cannot keep up with the throughput of current high-resolution EBCs. Therefore, a near-memory computing architecture is employed to accelerate the TOS update algorithm on the peripheral circuit. By utilizing an 8T SRAM with read-write decoupling, pipelining can be used resulting in a delay reduction of up to 24.7$\times$ compared to traditional cases. Additionally, DVFS is applied to dynamically adjust the operating voltage based on event rates, achieving 6.6$\times$ reduction in energy compared to traditional digital implementations. Using system level simulations of corner detection, we show that at the lowest power supply of 0.6 V, the decrease in AUC is only 0.027 and 0.015 for shapes\textunderscore dof and dynamic\textunderscore dof, respectively.

\vspace{12pt}

\bibliographystyle{IEEEtran}
\bibliography{ref}
\end{document}